\documentclass[structabstract]{aa}  
\usepackage{graphicx}
\usepackage{txfonts}
\usepackage{natbib}
\usepackage{float}
\usepackage{textcomp}
\usepackage{array}
\usepackage{placeins}
\DeclareTextSymbol{\degre}{OT1}{23}
\begin{document}
   \title{Population synthesis modelling of Luminous Infrared Galaxies at intermediate redshift}


   \author{E. Giovannoli
          \inst{1}
          \and
          V. Buat\inst{1}
          \and
          S. Noll\inst{2}
          \and
          D. Burgarella\inst{1}
          \and
          B. Magnelli\inst{3}
          }
          
   \institute{Laboratoire d'Astrophysique de Marseille, OAMP, Universit\'e Aix-Marseille, CNRS, 38 rue Fr\'ed\'eric Joliot-Curie, 13388 Marseille cedex 13, France\\
              \email{elodie.giovannoli@oamp.fr,veronique.buat@oamp.fr,denis.burgarella@oamp}
         \and
              Institut f\"ur Astro- und Teilchenphysik, Universit\"at Innsbruck, Technikerstr. 25/8, 6020 Innsbruck, Austria\\
               \email{Stefan.Noll@uibk.ac.at}
         \and      
             Max-Planck-Institut f\"{u}r Extraterrestrische Physik (MPE), Postfach
             1312, 85741 Garching, Germany.\\
             \email{magnelli@mpe.mpg.de}}


   \date{}
 
 
  \abstract  
  {Studying Luminous InfraRed Galaxies (LIRGs) is particularly important in the { build-up} of the stellar mass from z=1 to z=0, and to determine physical properties of these objects at the redshift 0.7. LIRGs are now identified to play a major role in galaxy evolution from z=1 to 0. 
  The global star formation rate (SFR) at $z \sim 0.7$ is mainly produced by LIRGs.}
   {  We perform a multiwavelengths study of a LIRGs sample in the Extended Chandra Deep Field South at z=0.7, selected at 24 $\mu$m by MIPS onboard $Spitzer~Space~Telescope$ and detected in 17 filters. Data go from the near-ultraviolet to the mid-infrared. This multiwavelengths dataset allows us to bring strong constraints on the spectral energy distributions (SEDs) of galaxies, and thus to efficiently derive physical parameters as the SFR, the total infrared luminosity, attenuation parameters and star formation history. We distinguish a sub-sample of galaxies detected at   70 $\mu $m which we compare to the rest of the sample to investigate the relative importance of this wavelength on the determination of the physical parameters. An important part of this work is the elaboration of a mock catalogue which allows us to have a reliability criteria for the derived parameters.}
   {We study LIRGs by means of a SED-fitting code CIGALE. At first, this code creates synthetic spectra from the Maraston (2005) stellar population models.
  The stellar population spectra are being attenuated by using a synthetic Calzetti-based attenuation law before the addition of the dust emission as given by the infrared SED library of Dale\&Helou (2002).
   The originality of CIGALE is that it allows us to perform consistent fits of the dust-affected ultaviolet-to-infrared wavelength range.
   This technique appears to be a very powerful tool in the case where we can have access to a dataset well-sampled over a large range of wavelengths. }
   {We are able to derive a star formation history and to estimate the fraction of infrared luminosity reprocessed by an active galactic nucleus. 
   We study the dust temperatures of our galaxies detected at 70 $\mu $m and find them colder than predicted by models. We also study the relation between the SFR and the stellar mass and do not find a tight correlation between both, but a flat distribution and a large scatter which is interpreted in terms of variations of star formation history.
   }
   { }
   \titlerunning{LIRGs at intermediate redshift}
   \maketitle
%

\section{Introduction}

Luminous InfraRed Galaxies (LIRGs) are commonly defined  as galaxies whose infrared  (IR, 8-1000 $\mu $m) emission is higher than $10^{11}$L$_\odot$ and lower than $10^{12}$L$_\odot$ . In a pionneering work \citet{CE01} (hereafter CE01), identified them as the main contributor to the cosmic IR background detected by $ISO$ ($Infrared~Space~Observatory$) ; the deep $Spitzer$/MIPS (\citet{rieke04})  surveys have confirmed that  they play a major role in galaxy evolution at intermediate redshift (z=0.5-1).
Indeed the IR luminosity function is found to evolve with z  and to shift towards higher luminosities when z increases (\citet{lefloch05}, \citet{perez-gonzalez05}) ; as a consequence,  LIRGs are  rare objects in the nearby universe but  become the  major contributors to the star formation density measured in IR  at z=0.5-1 (\citet{lefloch05}, \citet{caputi07}, \citet{magnelli09}).
The evolution of LIRGs as a function of z does not only concern their contribution to the luminosity density but also their physical characteristics. Most of the LIRGs populating the local universe are located in systems interacting violently (\citet{ishida04}). Conversely, at z $\sim$1 only 30$\%$ of LIRGs exhibit features linked to violent merging (\citet{bell07}, \citet{zheng07}) : most of them look like bright spirals who experience a secular evolution without violent events. This morphological difference between local and distant LIRGs is corroborated by the analysis of their star formation rate (SFR). Whereas local LIRGs are experiencing a strong starburst, distant LIRGs do not seem to strongly depart from the mean SFR - stellar mass (hereafter $M_{\star}$) relation found at z=1 (\citet{elbaz07}). 

Quite naturally, most of the studies devoted to LIRGs are based on IR data only. Very recently several teams aimed at studying local LIRGs by combining a large set of wavelengths and observations {(\citet{burgarella05},} \citet{kaviraj08}, {\citet{dacunha08}}). At higher z such studies  are scarcer. 
Zheng et al. (2007) have combined ultraviolet (UV), optical and IR observations and built mean spectral energy distribution (SEDs) of IR selected galaxies at z=0.7. 
They are found similar to those  of nearby normal galaxies with lower  IR luminosities (hereafter $L_{dust}$, $L(8-1000 \mu $m$)$).

The availability of large datasets of photometric data have lead to a renewal of stellar populations synthesis codes  assuming more or less complex star formation histories (SFHs). The combination of galaxy simulations with realistic SED modelling improves our understanding of uncertainties  of these fitting  methods  to estimate SFRs, $M_{\star}$ or population ages (\citet{wuyts09}, \citet{conroy09a}, \citet{lee09} , \citet{dye08}).  Anyway,  most of  these studies only deal with the stellar UV - Near-InfraRed (NIR) SEDs of galaxies. Since LIRGs emit most of their energy in IR we must account for the entire UV-to-IR SEDs to characterise them. Up to now there has been  few attempts to combine  stellar and dust emissions in the modelling of galaxy SEDs.  Stars and dust interact in a very complex way in galaxies and a  detailed description of this interplay is not always consistent with a SED-fitting code aimed at studying large samples of galaxies (\citet{dopita05}). The GRASIL (GRAphite and SILicate) code (\citet{silva98}, \citet{panuzzo05}) produces UV-to-IR SEDs based on a set of physical processes; \citet{iglesias-paramo07} used GRASIL to generate  a synthetic library of galaxy SEDs and fit a multiwavelengths dataset of UV selected galaxies from z=0.2 to 0.7. Da Cunha et al. (2008) developed a physically motivated model combining three IR  components to the stellar populations synthesis code of  \citet{BC03}.  Burgarella et al. (2005) and \citet{noll09} followed a  similar approach and proposed a code which calculates the effect of dust on galaxy SEDs in a consistent way; the code is versatile enough to be used at different redshifts and  with different SFHs. All these codes which combine stellar and dust populations are very efficient to constrain dust attenuation on the basis of an energetic budget : the IR dust emission originates from the stellar light absorbed in the UV-NIR part of the spectrum.
In this paper we  apply a multiwavelengths analysis from the far-ultraviolet (FUV) to the IR, based on SED-fitting, on a sample of z=0.7  LIRGs selected at 24 $\mu$m. Our analysis is based on the code CIGALE (Code Investigating GALaxy Emission \footnote{Web address to use CIGALE : http://www.oamp.fr/cigale/})  (\citet{noll09}).  It is the first application of this code to intermediate redshift galaxies mostly emitting in IR. 
Our aim is to study this galaxy sample, representative of LIRGs at intermediate redshift, in a very homogeneous and systematic way to determine the main characteristics of their stellar populations and dust emission.  We will  re-visit dependance of the dust temperature (deduced from the IR SED) on $L_{dust}$ and the SFH of these galaxies.

In section 2 we describe the sample and the multiwavelengths dataset. 
In section 3 we summarize the main characteristics of  the CIGALE code and its application to our sample, in particular we build  a mock catalogue to test the code performances. The results of the SED-fitting to the LIRGs sample are presented in section 4.  Section 5 is devoted to a discussion of these results in the framework of the  temperature-luminosity relation and the specific SFRs of LIRGs at z=0.7 which will be compared to model predictions.

We adopt an H$_0$=70 km s$^{-1}$ Mpc$^{-1}$, $\Omega_M$=0.3, $\Omega_{\Lambda}$ = 0.7 cosmology throughout this paper.
 

\section{The sample and data}
The Chandra Deep Field South (CDFS, \citet{giacconi02}) is one of the most intensely studied regions of the sky, with observations stretching from the X-ray to the radio, including UV, optical, IR, and submillimeter imaging, from space-based as well as the largest terrestrial observatories. 
The region of the CDFS ($\alpha$=3h32m00s, $\delta$=-27$\degre$~35'~00'', J2000.0) was observed with the MIPS instrument on board the $Spitzer~Space~Telescope$ (\citet{werner04}) in 2004 January as part of the MIPS Guaranteed Time Observing  programmeme. 

 \citet{lefloch05} used the  2955 sources detected at 24 $\mu$m by MIPS with flux at 24 $\mu$m $>$ 83$\;\mu$Jy  at the 80$\%$ completeness limit (\citet{papovich04}) to build the IR luminosity function from z=1 to 0. Photometric redshifts from COMBO-17 (Classifying Objects by Medium-Band Observations in 17 filters,  \citet{wolf04} ) were used for sources at z $\leqslant 1.2$ (see Le Floc'h et al. (2005) for more details). 
\citet{buat07-1} selected a subsample of 190 LIRGs at 0.6$<$z$<$0.8 and measured the near ultraviolet (NUV) emission of these galaxies on deep $Galaxy~Evolution~Explorer$ ($GALEX$) images. 80$\%$ of the 190 LIRGs were detected at NUV. 
The four IRAC bands (3.6$\mu$m, 4.5$\mu$m, 5.8$\mu$m and 8.0$\mu$m) are also available for these sources (N. Bavouzet, private communication). 
The optical-to-NIR photometry was taken from a K-selected catalogue of the ECDFS( Extended CDFS, $\alpha$=3h32m00s, $\delta$=-27$\degre$  48' 00'') as part of the multiwavelengths Survey by Yale-Chile (MUSYC, \citet{taylor09}  \footnote{MUSYC collaboration : http://www.astro.yale.edu/MUSYC/}). This catalogue consists of photometry derived from U, U38, B, V, R, I, z', J, H and K, imaging for approximately 80$\%$ of the field. The 5  $\sigma$ flux limit for point-sources is $K^{(AB)}_{tot}$ =22.0. 

We cross-correlate sources selected by \citet{buat07-1}  and detected at 24 $\mu $m with MUSYC data using a tolerance radius of 2".  When a double match is found, we select the closest object. We ignore the case where three or more optical sources can be associated with a given 24 $\mu$m detection and thus eliminate 9 objects from the 190 sources. {At the end we are left with 181 galaxies whose 150 have a NUV detection (Table~\ref{table1}) with a mean redshift of 0.70 +/-0.05. For the 31 objects not detected in NUV a mean redshift of 0.72 +/- 0.05 is found.}

The ECDFS was observed with the MIPS instrument as part of the FIDEL (Far-Infrared Deep Extragalactic Legacy Survey, PI : Dickinson)  legacy programmeme . For the 70 $\mu $m data, we used the catalogue from \citet{magnelli09} . Detection limit for 70  $\mu $m sources in this field is 3.5 mJy (5 $\sigma$).
We refer to this paper for more details about the source extraction, photometry and completeness of the catalogue.
We cross-correlated our 181 objects with FIDEL data thanks to their IRAC ( \citet{fazio04}) positions, with a  a tolerance radius of 2". Finally 62 galaxies were detected at 70 $\mu$m {(Table~\ref{table1}), with a mean redshift of 0.68 +/- 0.04.}

The fraction of MIPS sources showing evidence for the presence of an Active Galactic Nucleus (AGN) in their optical counterparts is less than $\sim$15$\%$ according to the VIMOS (VLT Deep Survey) and COMBO17 classifications. Synthetic models connecting the X-ray and IR SED of AGN  indicate that the emission arising from pure AGNs should be negligible (i.e., $\lesssim$ 10$\%$) in high redshift sources detected at 24 $\mu $m ( \citet{silva04}).  To have more precision about the possible fraction of AGN in our sample we use a diagnostics colour-colour given by  \citet{stern05} and the diagnostics  of the mid-IR (MIR) slope of \citet{brand06} (see section 3.3) ; we also use a specific output of the code CIGALE (see section 4.2). 
These studies will be described later in this paper.

We decide to distinguish our SED-fitting analysis between the sub-sample detected at   70 $\mu $m and the whole sample. Hereafter we will use ' 70 $\mu $m  sample' for galaxies detected at   70 $\mu $m and 'total sample' for the whole set of galaxies.

\begin{table*}
\centering
\begin{tabular}{lccc}
 \hline
  \textbf{Filters}&\textbf{Wavelength}& \textbf{Instrument-Survey} & \textbf{Galaxies detected}  \\
  \hline
   \textbf{Mid-IR, IR} & & & \\
  \hline  
   MIPS 1& 24 $\mu$m & MIPS/Spitzer - FIDEL & 181\\
   MIPS 2 &   70 $\mu $m & " & 62\\
  \hline  
    \textbf{Near-IR} & &  \\
  \hline  
 IRAC 1 & 3.6 $\mu$m & IRAC-GTO Spitzer & 177\\
  IRAC 2 &4.5$\mu$m & " & 161 \\
  IRAC 3 &5.8$\mu$m & " & 177 \\
  IRAC 4 &8.0$\mu$m & "  & 161\\
   \hline
  \textbf{UV - Optical} & & &\\
  \hline  
 U  &  0.35 $\mu$m& MUSYC & 179\\
 U38 &  0.36 $\mu$m& " & 180\\
 B &  0.46 $\mu$m& "  & 181\\
 V &  0.54 $\mu$m& " & 181\\
 R &  0.65 $\mu$m& " & 181\\
 I &  0.87 $\mu$m& " & 181\\
 z &  0.90 $\mu$m& " & 181\\
 J &  1.2 $\mu$m& " & 180 \\
 H &  1.6 $\mu$m& "& 155\\
 K &  2.1 $\mu$m& " & 181\\
   \hline
  \textbf{UV} \\
          \hline  
 NUV & 2310 \AA & GALEX-DIS  &  150\\
   \hline
  \end{tabular}
     \caption{The observed data : wavelengths corresponding to the filters identification of instrument and survey. Our sample contains 181 galaxies of which 62 are detected at 70 $\mu$m.}
   \label{table1}
\end{table*}


\section{ SED-fitting techniques : description of the code CIGALE}

The best way to derive physical parameters like the SFH is to fit the observed SED with models from a stellar populations synthesis code.
We use the code CIGALE which provides physical informations about galaxies by fitting their UV-to-IR SED. 
A first version of the code was developed by \citet{burgarella05} to reproduce the dust-attenuated stellar emission from the FUV to the NIR. 
$L_{dust}$ could be provided as an input parameter only. Hence, \citet{noll09}  especially extended the code to the far-IR (FIR) by the consideration of dust emission models to allow a consistent treatment of dust effects on galaxy SEDs.
A Bayesian-like analysis is used to derive galaxies properties and  \citet{noll09} applied it to local galaxies. This work represents the first attempt to apply CIGALE on galaxies at intermediate redshift.
CIGALE is based on the use of a UV-optical stellar SED plus a dust IR emitting component. First , UV-to-IR models are built and secondly, these models are fit to the observed SEDs. Practically the code fits the observed data in the UV, optical and NIR with models generated with a stellar populations synthesis code, assuming a SFH and a dust attenuation scenario as input. 
{The energetic balance between dust-enshrouded stellar emission and re-emission in the IR is carefully conserved in combining the UV/optical and IR SEDs. }
The IR SEDs are built from the  \citet{DH02} templates (hereafter DH02).  We refer the reader to \citet{noll09} for more details. 

In the following we will describe parameters crucial to this study.
The input parameters values are listed in Table \ref{table4} . 

\subsection{Stellar populations and star formation histories}

In its current version CIGALE provides two stellar populations synthesis models : Maraston et al. (2005) (hereafter M05) and PEGASE ( \citet{pegase97}). Both include a treatment of thermally pulsating asymptotic giant branch (TP-AGB) stars but in a different way.
For instance, in M05 models young stellar populations are modelled with the Geneva stellar evolutionary tracks, while the Padova tracks are used in PEGASE.  

TP-AGBs are particular important for a reliable $M_{\star}$ determination at high z (M05). 
The difference between PEGASE and M05 models lies in the weight given to the TP-AGBs. Their contribution will be systematically higher in M05 models. Models which do not consider enough TP-AGB  need a massive contribution of old stars in order to match the observed NIR fluxes and thus overestimate $M_{\star}$ (M05).
In a recent work \citet{tonini09} found that the use of PEGASE does not reproduce IRAC luminosities for nearly passive and star-forming galaxies around z $\sim$ 2.
For these reasons we decide to use the M05 models.

 
 Two stellar components have been found necessary to reconstruct more accurate SFRs in actively star forming galaxies (e.g \citet{erb06,lee09}).
CIGALE uses the single stellar populations of M05 and combines two stellar populations to reproduce an old and a young stellar population.
M05 models assume SFH with exponentially decreasing SFR : $\tau_{1}$ and $\tau_{2}$ represent the e-folding time of the old and young stellar population, respectively.  
The SFR is expressed as follow: 

\begin{equation} 
\mathrm{SFR}_ {1}(t) = \mathrm{SFR}_{0, 1}.e^{\frac{-(t-t_1)}{\tau_{1}}}  
 \end{equation}
 \begin{equation}
\mathrm{SFR}_ {2}(t) = \mathrm{SFR}_{0, 2}.e^{\frac{-(t-t_2)}{\tau_{2}}}  
\end{equation}
 $t_1$ and $t_2$ represent the age of the old and the young stellar population  and SFR$_{0, 1}$ and SFR$_{0, 2}$ are the values of the SFR at t=t$_1$ and t=t$_2$, respectively. 

We study a sample of LIRGs lying at z=0.7. At this redshift the universe was 7 Gyr old with our cosmology, so it makes no sense to consider a range up to more than 7 Gyr for the ages of both stellar populations.
 After several tests we realized that we cannot make a precise estimate of the e-folding time of the young stellar population, so we only consider a SFR constant over the time. 
We decide to fix the age of the old stellar population $t_1$=7 Gyr and we suppose 3 possibilities for  $\tau_1$ : $\tau_1$ = 1,3 or 10 Gyr. We choose $t_2$ in the range 0.025 to 2 Gyr and we consider $\tau_2$ = 20 Gyr which corresponds to a burst constant over this time-scale.
 
 The two SFH components are linked by their mass fraction. The fraction of the young stellar population (hereafter $f_{ySP}$) corresponds to the fraction of the young stellar population mass over the total mass and is thus comprised between 0 and 1. We chose values of  $f_{ySP}$ in this interval to allow the choice of the best fitting value, as we do not have any prescription for this parameter.
{CIGALE gives an instantaneous value of  log$_{10}$SFR as an output, defined as log [(1-$f_{ySP}$)*SFR$_1$ + $f_{ySP}$*SFR$_2$]}.{ In the following, any occurrence to the SFR will refer to this formula.
} 
  \begin{figure}
\centering
 \includegraphics[width=9cm]{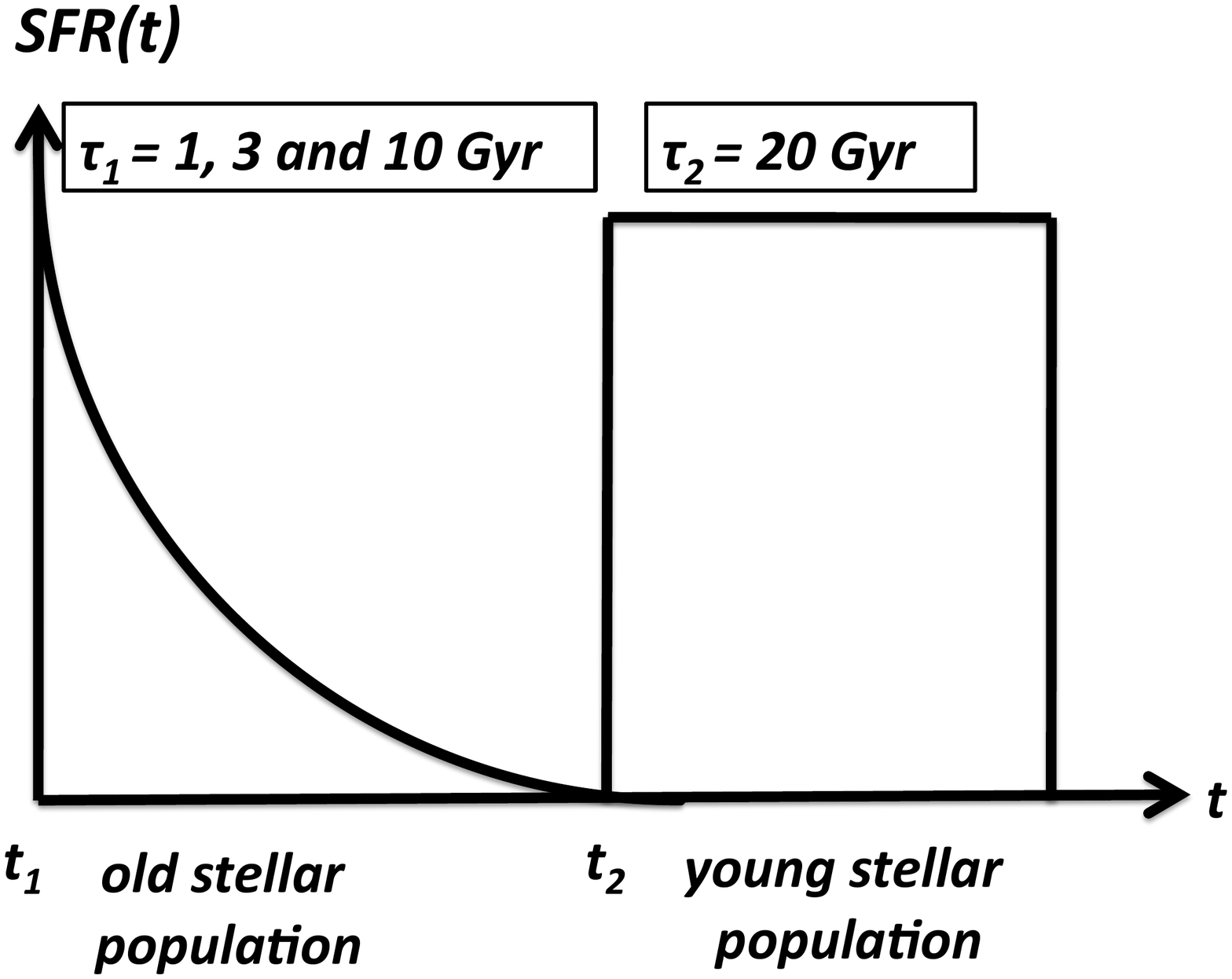}
    \caption{Scenario of SFH  : two bursts, one for the old stellar population having a moderate e-folding time and one for the young stellar population constant over time.}
    \label{SFH_2}    
\end{figure}
 
\subsection{Dust attenuation and infrared emission}
To modelise the attenuation by dust, the code CIGALE uses attenuation law of \citet{calzetti00} as a baseline, and offers the possibility to vary the attenuation law and to add a bump.
We decide to reproduce a pure Calzetti law and to add no bump.
The stellar population models have to be attenuated before the IR emission can be added depending on the dust-absorbed luminosity.
The code allows to consider that young and old stars are not affected by dust in the same way.
Dust attenuation in the V band, ${A_V}$, and the dust reduction factor $f_{V}$ are used for the young and the old stellar population respectively.  
The stellar light heats the dust which reemits the energy in MIR and FIR. At these wavelengths we have few available data to reconstruct the SED. To fit IR observations CIGALE uses semi-empirical one-parameter models of DH02 composed of 64 templates. These models are parametrised by $\alpha$, the power law slope of the dust mass over heating intensity, defined as follows :

\begin{equation}
dM_ {d}(U)  = U^{\alpha}dU  \\
 \end{equation}
where $M_d(U)$ is the dust mass heated by a radiation field at intensity U ;{ $\alpha$ is directly related to $f_{60\mu m}$/$f_{100 \mu m}$ ratio flux, where $f_{60\mu m}$ and $f_{100\mu m}$ represent fluxes of the SED at 60 and 100  $\mu$m, respectively.}

 DH02 provide SEDs corresponding to different exponents $\alpha$, which is a free parameter only for the   70 $\mu $m sample. 
The total sample only has 24 $\mu$m fluxes, so we take the median value of $\alpha$ found by the code for the 70 $\mu$m sub-sample. 
  We choose $\alpha$ in the interval [1 ; 2.5]  to cover a large domain of activity (the activity increases as $\alpha$ decreases).
\subsection{Active galactic nuclei contribution}

LIRGS have large $L_{dust}$ corresponding to an active phase of dust enshrouded star formation and/or AGN activity. 
The energetic balance between dust-enshrouded stellar emission and re-emission in the IR can be violated if dust emission caused by an AGN is present. 
In particular, highly dust-eshrouded AGNs represent a problem since they are difficult to identify in the UV and optical. Consequently, galaxies with such an AGN contribution may look like normal star-forming galaxies in the FUV-to-NIR wavelength range excepting a warm/hot dust emission in the MIR.

In this paper we are interested in the $M_{\star}$ building by LIRGs.
Therefore it is essential to separate the contribution of AGN and starbursts to $L_{dust}$.
The AGN contribution is likely to be dominant at only very high luminosity ($L_{dust}\gtrsim 2.10^{12}$ L$_{\odot}$, {\citet{sandersmirabel96}}). Nevertheless, a small fraction of the total  IR luminosity can be due to the presence of an AGN even in star-formation dominated cases (e.g.  \citet{genzel98}). 

To disentangle IR dust emission components caused by stellar and AGN radiation, we apply the diagnostics of \citet{stern05}  to our observations. They combine spectroscopic observations with MIR observations of nearly 10000 sources at 0 $<$ z $<$ 2 and show that MIR photometry provides a robust technique for especially identifying broad-line AGN (90$\%$ of broad-line AGN and 40$\%$ of narrow-line AGN are identified).

\begin{figure}
\centering
 \includegraphics[width=9cm]{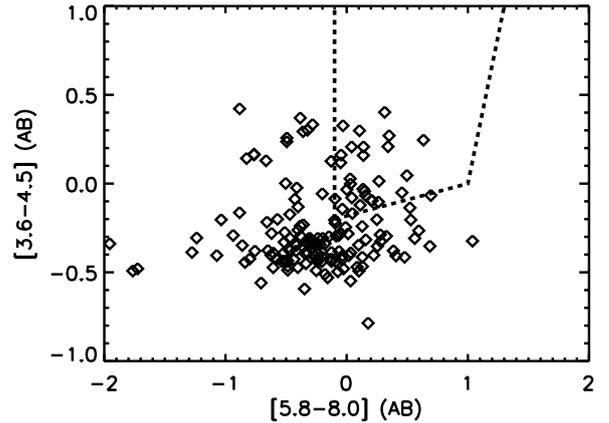}
    \caption{Color-color diagram preconized by \citet{stern05}  to identify AGN candidates. We compare the color [3.6-4.5] versus [5.8-8.0] in AB magnitudes for our LIRGs; the area defined by the dashed lines represents the region of the color-color diagram populated by AGN. 
}
    \label{AGN}    
\end{figure}

Fig. \ref{AGN} shows the colour-colour distribution proposed by \citet{stern05} in the NIR. The dashed lines represent the boundary of the area which characterise AGN features. 15 $\%$ of the objects  are found  within this area. However these colour criteria may omit AGNs at z  $\approx$ 0.8 and 2 ( \citet{stern05}).

\begin{figure}
\centering
 \includegraphics[width=9cm]{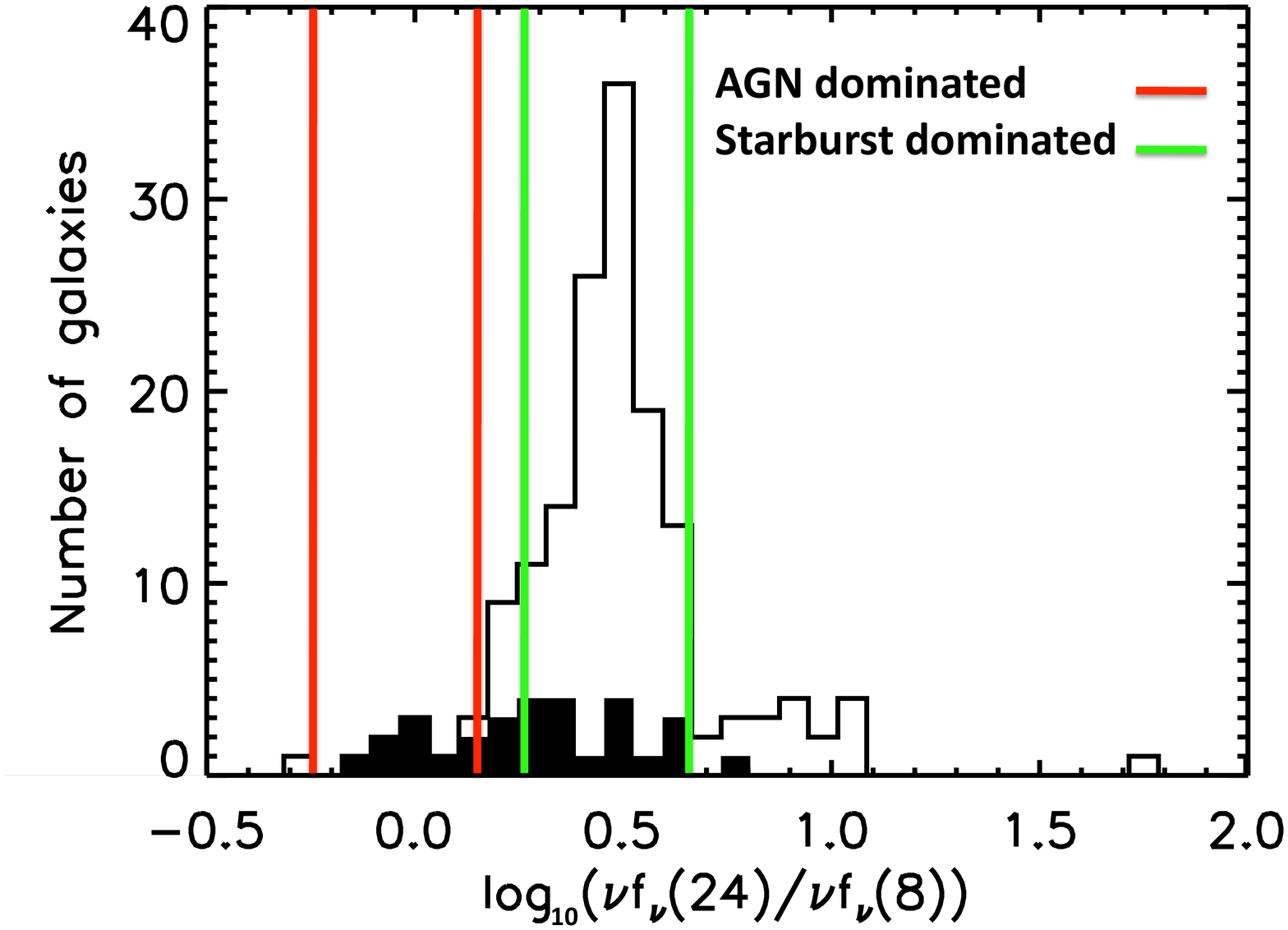}
    \caption{The empty histogram illustrates the flux ratio  $\zetaup$ = log$_{10}(\nu f_{ \nu}(24~ \mu m)/\nu f_{ \nu}(8 ~\mu m))$  for the total sample. 
     The black histogram shows the galaxies identified as AGNs by the diagnostic of Stern et al. (2005).
     }
    \label{AGN_brand}    
\end{figure}

We also use the diagnostics of \citet{brand06} who examinate the MIR slope $\zetaup$ of the SEDs for their sample. $\zetaup$ corresponds to the observed flux ratio log$_{10}(\nu f_{ \nu}(24~ \mu$m$)/\nu f_{ \nu}(8 ~\mu $m$))$.  
The MIR slope is supposed to be steeper for starburst-dominated sources.
 \citet{brand06} assume that sources (z$>$0.6) with $\zetaup$ peaking at $\approx$ 0.0 and $\approx$ 0.5 are AGN- and starburst-dominated, respectively. Fig.~\ref{AGN_brand} shows the flux ratio  $\zetaup$  for the total sample. 
For the vast majority of objects $\zetaup$ is found around 0.5 ($\sim$17$\%$ satisfies  $\zetaup$ $\approx$ 0.0) which lets us think that starbursts dominate the MIR emission of our sample. 
We overplot in black the histogram of the objects previously identified as AGNs by Stern et al. (2005). We find that they are not necessarily identified by Brand et al. (2006).

\begin{figure}
\centering
 \includegraphics[width=8cm]{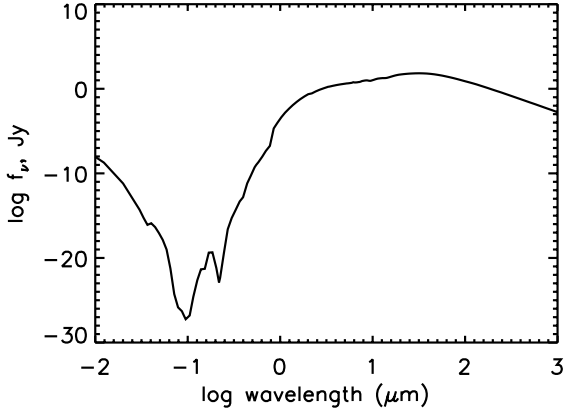}
    \caption{{ SED of the adopted AGN template, with  $L = 10^{12} $L$_{\odot}$, $R=125$ pc, and $A_V = 32$ mag}.
     }
    \label{AGN_template}    
\end{figure}

CIGALE allows to estimate the fraction of  $L_{dust}$ due to an AGN.
 The code uses AGN templates from \citet{siebenmorgen04a,siebenmorgen04b} who provide almost 1500 AGN models differing in the luminosity of the non-thermal source, the outer radius of a spherical dust cloud covering the AGN, and the amount of attenuation in the visual caused by clouds.
All models can be fed into CIGALE. Focusing on SEDs providing PAH-free MIR emission, the number of suitable models signiÞcantly decreases. As the reference model, we take $L = 10^{12} $L$_{\odot}$, $R=125$ pc, and $A_V = 32$ mag { (see Fig.~\ref{AGN_template} ).} The parameter in our runs which probes the contribution of AGN radiation to $L_{dust}$ is named  $f_{AGN}$ for fraction of AGN .
As input we propose values for $f_{AGN}$ ranging from 0.0 to 0.999. The results will be presented in section 4.2.

Later we will decide to flag objects  identified as AGN by at least 2 diagnostics  ( \citet{stern05} ,  \citet{brand06}, CIGALE).

\begin{table*}
 \centering
\begin{tabular}{ l c l}

  \hline
  {Parameters} & {Symbol} & {Range} \\
  \hline
  {Attenuation} & &  \\
  \hline  
  V-band attenuation & $A_{V}$ & 0.15; 0.30; 0.45; 0.60; 0.75; 0.90; 1.05; 1.20; 1.35;  \\
                                       &                &1.5; 1.65; 1.8; 1.95; 2.1 \\
  Reduction of $A_{V}$  basic for old SP model & $f_V$ & 0.0; 0.25; 0.50; 0.75; 1.0 \\
    \hline                                   
 {IR SED} & &\\
    \hline
 IR power-law slope & $\alpha$ & 1.0 ; 1.25; 1.5; 1.75; 2.0; 2.25; 2.5  \\
 AGN-related fraction of $L_{dust}$, & $f_{AGN}$ & 0.0; 0.001; 0.005; 0.01; 0.05; 0.1; 0.5; 0.7; 0.999  \\
  \hline
 {Star Formation History} & &   \\
  \hline   
metallicities (solar metallicity) & $Z$ &  0.02 \\
 $\tau$ of old stellar population models in Gyr & $\tau_1$ & 0.1;  3.0; 10.0 \\ 
ages of old stellar population models in Gyr & $t_1$ & 7  \\
$\tau$ of young stellar population models in Gyr & $\tau_2$ & 20.0 \\ 
ages of young stellar population models in Gyr & $t_2$ & 0.025; 0.05; 0.075; 0.1; 0.2; 0.3; 0.4; 0.5; 0.6; 0.7;  \\
                                                                                     &             & 0.8; 0.9; 1.0; 1.5; 2.0 \\
fraction of young stellar population & $f_{ySP}$ & 0.0; 0.001; 0.005; 0.01; 0.05; 0.06; 0.07; 0.08; 0.09; \\
                             &                    & 0.1; 0.2; 0.3; 0.4; 0.5; 0.6; 0.7; 0.8; 0.999 \\
IMF & S & Salpeter  \\
  \hline   
  \end{tabular}
  \caption{List of the input parameters of the code CIGALE and their selected range for a specific run.}
  \label{table4}
\end{table*}

\subsection{Do we retrieve reliable parameters? Construction of a mock catalogue dedicated to the sample.}

The code gives an estimate of the basic input parameters plus several additionnal output parameters that depend on different basic model properties. 
We mostly focus on SFH parameters for the young stellar population.

We perform a Bayesian analysis to estimate the physical parameters. We use a method similar to the one of \citet{kauffmann03a}. They derive a probability distribution function (PDF) for each parameter and estimate the most probable value. We use the mean value of this distribution. The expectation value of each parameter $ <x>$ is given as   : 
 
 \begin{equation}
 <x>=  \frac{ \sum_{i=1}^b P_ix_i}{ \sum_{i=1}^b P_i}
 \end{equation}
and the standard deviation $ \sigma$ is derived by :

 \begin{equation}
 \sigma_x=  \sqrt{\frac{ \sum_{i=1}^b P_i(x_i-<x>)}{ \sum_{i=1}^b P_i}}
 \end{equation}
 This method is  described as the 'sum' method in \citet{noll09}.

When we use a numerical code based on SED-fitting, we obtain results depending on a statistical analysis. The accuracy of the estimates is directly dependent on the input fluxes (number and quality).
We must check which parameter the code is able to estimate correctly.
Theoretically the Bayesian analysis provides an estimation of the reliability of each parameter thanks to the shape of their PDF (if the shape is not a Gaussian curve the parameter is not well determined (\citet{walcher08}), and their relative error  $\sigma_x$. 

A more straightforward approach to check the parameter estimation is to generate a mock catalogue from the real sample.
Our general strategy is to build a specific mock catalogue for each set of data (in this paper the LIRGs sample). 
It is made of artificial galaxies for which we do not only know the flux at each wavelength but also values of attenuation parameters and parameters of the SFH.

To construct our mock catalogue we follow 3 steps. The first step is to run the code on the data to determine the best fitting model for each object by a simple {$\chi^2$} minimization method. Each corresponds to a SED. 
In the second step we add an error randomly distributed according to a Gaussian curve with $\sigma$=0.1 to each flux ; it means that we add an error to the flux which is typically 10$\%$ of the flux. 
We have a catalogue of artificial galaxies, detected in the same filters as the observed galaxies. Then the last step is to run the code on these simulated data and to compare the exact values of the physical parameters with the values estimated by the code.

\section{ Results of the SED-fitting analysis}

We perform SED-fitting on our LIRGs sample and on the related  mock catalogue in order to know which parameter we are able to correctly estimate. In the following sections we will describe the results of the SED-fitting of the mock catalogue with and without the band at   70 $\mu $m and then we will discuss results of the SED-fitting of the real data.

\subsection{Analysis of the mock catalogue}

We construct  two mock catalogues as described in { Sect. 3.4} based on the   70 $\mu $m sample (sample detected at 70$\mu$m) only. 
The first catalogue is made of galaxies detected at   70 $\mu $m  ( 70 $\mu $m  mock sample) and the second one is made with the same galaxies for which we remove the flux at   70 $\mu $m  (mock sample) in order to know if the fact of having   70 $\mu $m flux has an influence on the determination of the parameters of the SFH, or $M_{\star}$, $L_{dust}$ and SFR. 

\begin{figure}
\centering
 \includegraphics[width=9cm]{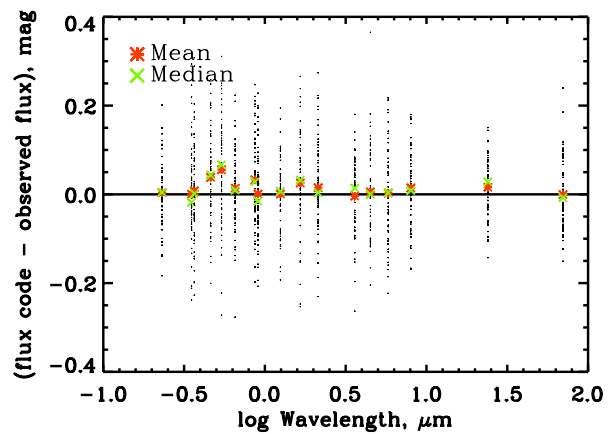}
    \caption{ Difference in magnitudes between the flux of the mock galaxies and the flux of the best $\chi^2$ model for these objects in each band. Red crosses and green crosses represent the mean and the median value for each wavelength.}
    \label{IO_fluxes_70}    
\end{figure}

In Fig. ~\ref{IO_fluxes_70} is plotted the difference in magnitudes between the flux from the best $\chi^2$ model estimated by CIGALE and the observed flux for each broad band filter.  Mean and median values are nearly equal to zero which means that fluxes are well reproduced by the code to better than 0.1 mag. 

\begin{figure*}
\centering
 \includegraphics[width=18cm]{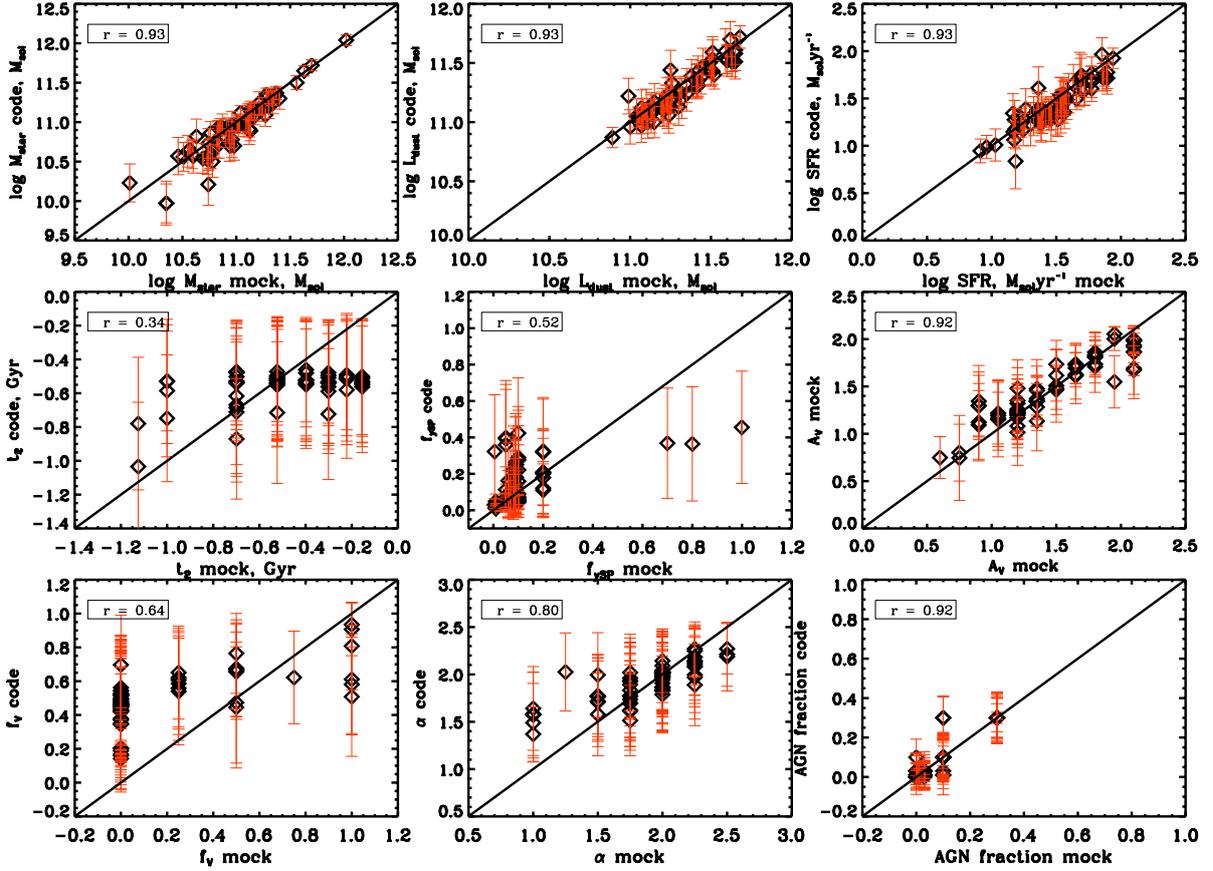}
    \caption{ On the x axis the value for 8 parameters of the mock galaxies in the case of a detection at 70 $\mu$m, and on the y axis their values estimated by CIGALE.  The red line corresponds to the ratio 1:1. }
    \label{probabay_70}    
\end{figure*}

\begin{figure*}
\centering
 \includegraphics[width=18cm]{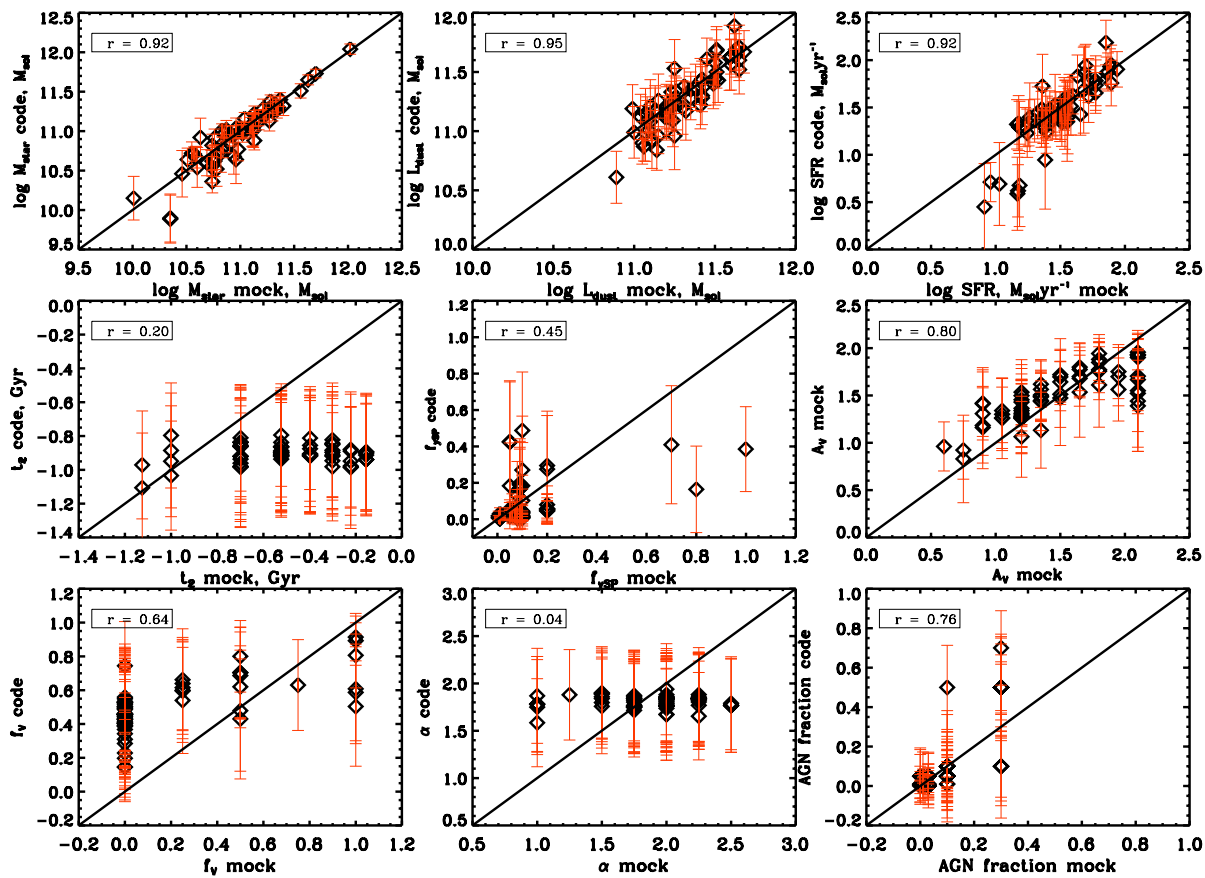}
     \caption{ Same plot as Fig.~\ref{probabay_70} in the case of non-detection at 70 $\mu$m.}
    \label{probabay}    
\end{figure*}

Even if the code is able to well reproduce the observed data, we cannot settle for this result. Indeed, it is possible that some SEDs are degenerated and correspond to different parameter values. We have to compare each input parameter with its Bayesian estimation by the code. Fig. ~\ref{probabay_70} and ~\ref{probabay}  show the input parameters of the mock galaxies  versus the values estimated by the code.  Fig. ~\ref{probabay_70} (Fig. ~\ref{probabay}) shows results for the 70 $\mu $m mock sample (mock sample).
Values of the Linear Pearson correlation coefficient r are gathered in Table ~\ref{table5}.

For the 70 $\mu $m mock sample we find very good correlations for  $M_{star}$, SFR, $L_{dust}$, $f_{AGN}$, and $A_V$ with r $>$ 0.9.
For $t_2$,  $f_{ySP}$ and $f_V$ the correlation is less satisfying, r is equal to 0.34, 0.52, and 0.64 for the 70 $\mu $m  mock sample and 0.20, 0.45, and 0.64 for the mock sample.
 The correlation for the parameter $\alpha$ is good with r=0.80 for    70 $\mu $m  mock sample but very bad for the mock sample.
In the following we will use the mean value of $\alpha$ from the 70\,$\mu$m sample for the objects without detection at 70 $ \mu$m. 
For $f_{AGN}$ r $>$ 0.9 for 70 $\mu $m mock sample and 0.76 for the mock sample, with relatively small errors. Thus this parameter is correctly estimated by the code. 

According to the analysis of the coefficient r and the Bayesian method's error, the reliable parameters are : $M_{star}$, $L_{dust}$, SFR, $f_{ySP}$, $A_V$, $\alpha$ (if   70 $\mu $m is available) and $f_{AGN}$. Consequently in the next section we will only study these parameters.

\begin{table}[h!b!p!]
\centering
\begin{tabular}{|l|l|l|}
  \hline
 \textbf{Parameters}  & \textbf{70 $\mu$m mock sample} & \textbf{ mock sample} \\
  \hline
log$_{10}~M_{star}$ &   0.93   & 0.92  \\
log$_{10}~L_{dust}$   &   0.93   &  0.95 \\
log$_{10}~SFR$ & 0.93 & 0.92\\
$t_2$ & 0.34 & 0.20 \\
 $f_{ySP}$ & 0.52& 0.45 \\
 $A_V$ & 0.92 & 0.80 \\
 $f_V$ & 0.64  & 0.64 \\
  $\alpha$ & 0.80 &  0.04 \\
  $f_{AGN}$ & 0.92 & 0.76  \\
  \hline  
  \end{tabular}
  \caption{Estimation of the linear correlation coefficient of Pearson between the exact value and the value estimated by CIGALE for  each parameter of the mock catalogue. We  estimate this coefficient for the 70\,$\mu$m mock sample and the mock sample.}
  \label{table5}
\end{table}


\subsection{Analysis of the LIRGs sample}

\begin{figure*}
\centering
 \includegraphics[width=20cm]{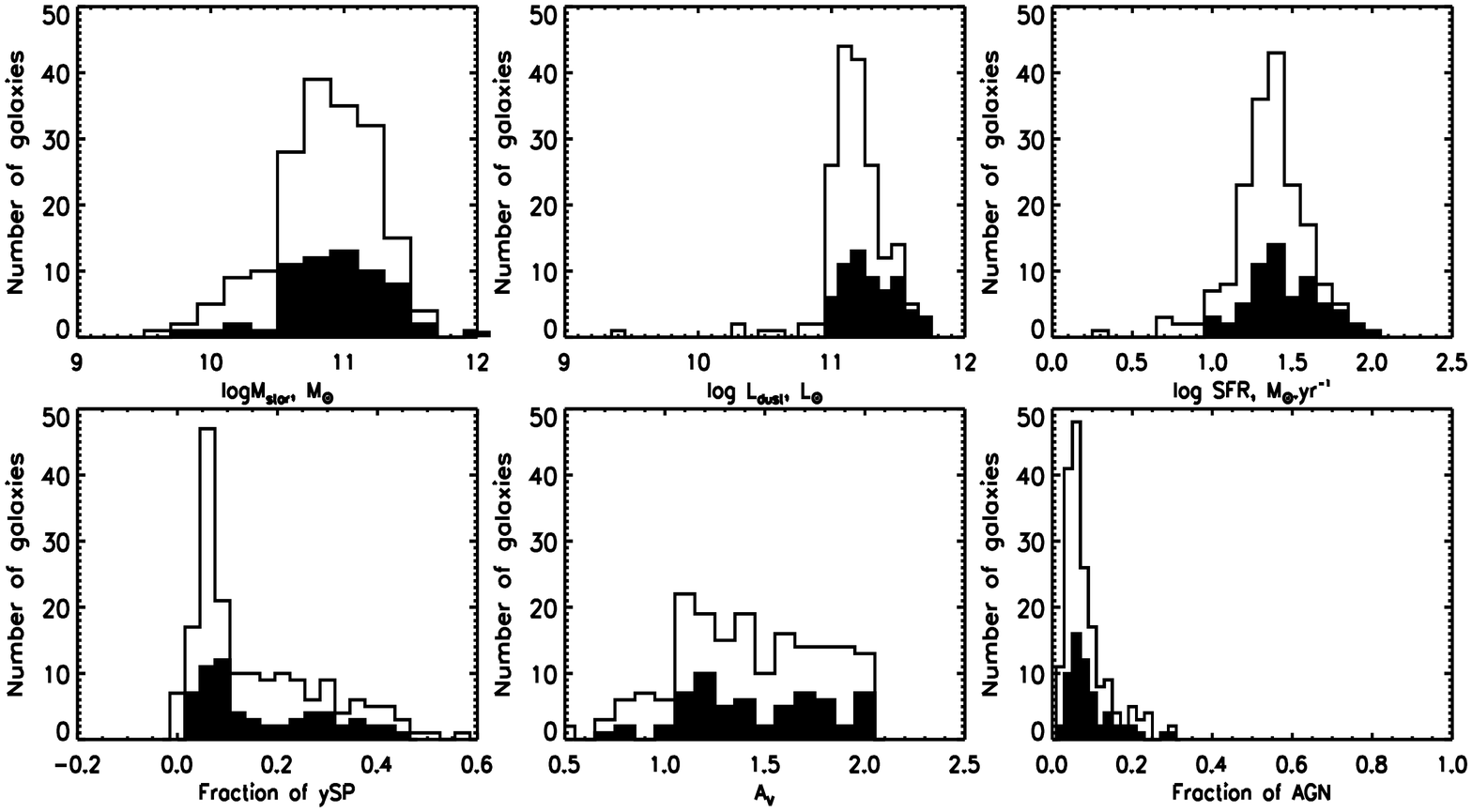}
    \caption{Bayesian results of the code for the following parameters : $M_{star}$, $L_{dust}$, SFR, $A_{V}$, {$f_{ySP}$} and $f_{AGN}$. The empty histogram represents the total sample and the full one represents the   70 $\mu $m sample. }
     \label{HISTO}
\end{figure*}

We now present the results for the LIRGs sample itself. Fig.~\ref{HISTO} shows  distributions for parameters of interest, calculated with the Bayesian analysis described previously for the sub-sample detected at   70 $\mu $m  and for the whole sample. 

The masses found for  the 70 $\mu $m sample are shifted towards larger masses in comparison with the total sample. We observe a similiar situation for $ L_{dust}$ and the SFR; the values are in the range  [ $10^{11}$ ; $10^{12}$] L$_{\odot}$ and [10 ; 92] $\rm{ M_{\odot}.yr^{-1}}$ respectively with mean values larger than ones found for the total sample. This shift is expected because of the 70 $\mu $m detection limit (2.7 mJy at 5$\sigma$) : at this wavelength we detect only luminous and massive galaxies. 
The distribution of  $f_{ySP}$  is broad with a long tail towards high values.
$A_V$ lies between 0.5 and 2.1 mag with very few objects under 1.0 mag and a quite homogeneous distribution between 1.0 and 2.0.  

Galaxies in the total sample have $M_{\star}$ between $10^{10}$ and $10^{12}$ M$_{\odot}$ with a peak at  $10^{10.8}$M$_{\odot}$. We find the SFR between 3 and 92 $\mathrm{M}_{\odot} \mathrm{yr}^{-1}$ with a peak at  23  $\mathrm{M}_{\odot} \mathrm{yr}^{-1}$.
For $f_{ySP}$, and $A_V$ we observe the same distribution as for the   70 $\mu $m sample.
For both samples $f_{AGN}$ is relatively low, between 0.0 and 0.3 with the majority of the objects in the interval [0;0.1]. We consider that there is a real contamination of $L_{dust}$ by an AGN {when} $f_{AGN}$ $>$ 15$\%$ because a contamination lower than15$\%$ does not significantly modify the total IR emission.
We must also account for the uncertainty of the $f_{AGN}$  determination : we only consider as AGN contaminated objects with $f_{AGN}$$>$ 15$\%$  and  $f_{AGN}$-$\sigma_{AGN}$$>$ 0 where $\sigma_{AGN}$ is the standard deviation (see Eq. 5).

We identify 21 objects in the total sample.
The diagnostics of \citet{stern05} identifies 31 objects, and \citet{brand06} 9 objects.
However, all the objects considered as AGN by the code are not systematically identified as AGN in the diagnostics of \citet{brand06} and \citet{stern05}. 
We consider objects as contaminated by AGN when they are flagged by two of the three diagnostics. We identify 15 objects  in the total sample of which 6 are in the 70 $\mu $m sample ;   these objects will be flagged in the following plots.                 

{We have also investigated the small subsample of galaxies not detected in NUV (31 over 181 galaxies) and did not find any statistically significant difference in their main characteristics. The largest effect is found for A$_{V}$ : 
A$_{V}$=1.75 +/- 0.25 mag for the non detections in NUV and A$_{V}$=1.33+/- 0.34 mag for the detections in NUV.
A higher dust attenuation for galaxies not detected in NUV is indeed expected.}


\section{Discussion}

\subsection{Luminosity-colour distribution and the implication for the dust temperature}


SED shape of a LIRG in the 8-1000 $\mu$m range is due to emission from dust.
The SED rest-frame peak defines the wavelength of maximum energetic output (in $\nu f_{\nu}$) ; it is usually in the 40-140 $\mu$m range and the SED is often approximated by a modified black-body ($B_{\nu} \nu^{\beta}$) of T $\sim$ 20-60 K and emissivity $\beta \sim$ 1-2.

Local $IRAS$ ($Infrared~Astronomical~Satellite$) galaxies exhibit correlations between the color $f_{60\mu m}$/$f_{100\mu m}$ and $L_{dust}$ (e.g \citet{dale01}).  $f_{60\mu m}$/$f_{100\mu m}$ ratio is often taken as an indicator of the typical heating conditions in the interstellar medium and is indicative of a characteristic dust temperature (hereafter $T_{dust}$) {, $f_{60\mu m}$/$f_{100\mu m}$ increasing with $T_{dust}$.}

In the nearby universe LIRGs are characterized by emission from dust at high temperatures (\citet{sandersmirabel96}).   Fitting single $T_{dust}$ to $f_{60\mu m}$/$f_{100\mu m}$ ratio leads to $T_{dust} \sim$20K for low redshift spirals (\citet{reach95}; \citet{DE01}) and 30-60K for the high-luminosity objects typically detected by $IRAS$ (\citet{SN91} ; \citet{standford00}). High-redshift, hyperluminous galaxies can show $T_{dust}$ of up to 110K (\citet{lewis98}).
 
 To reproduce number counts at FIR wavelengths, \citet{lagache03}, \citet{chapin09}, and \citet{valiante09} found that it is necessary to take into account cold luminous galaxies in the nearby universe.
Cold galaxies are found to dominate the z=0 luminosity function at low luminosities and become less important at higher redshifts, because of their passive evolution. The fraction of cold galaxies can contribute to the number counts up to 50$\%$ at  170 $\mu$m. 
 At moderate and high redshift the possible importance of cold, luminous galaxies has also been pointed out by \citet{eales99-00} and  \citet{chapman02c}.


\begin{figure} 
 \includegraphics[width=9cm]{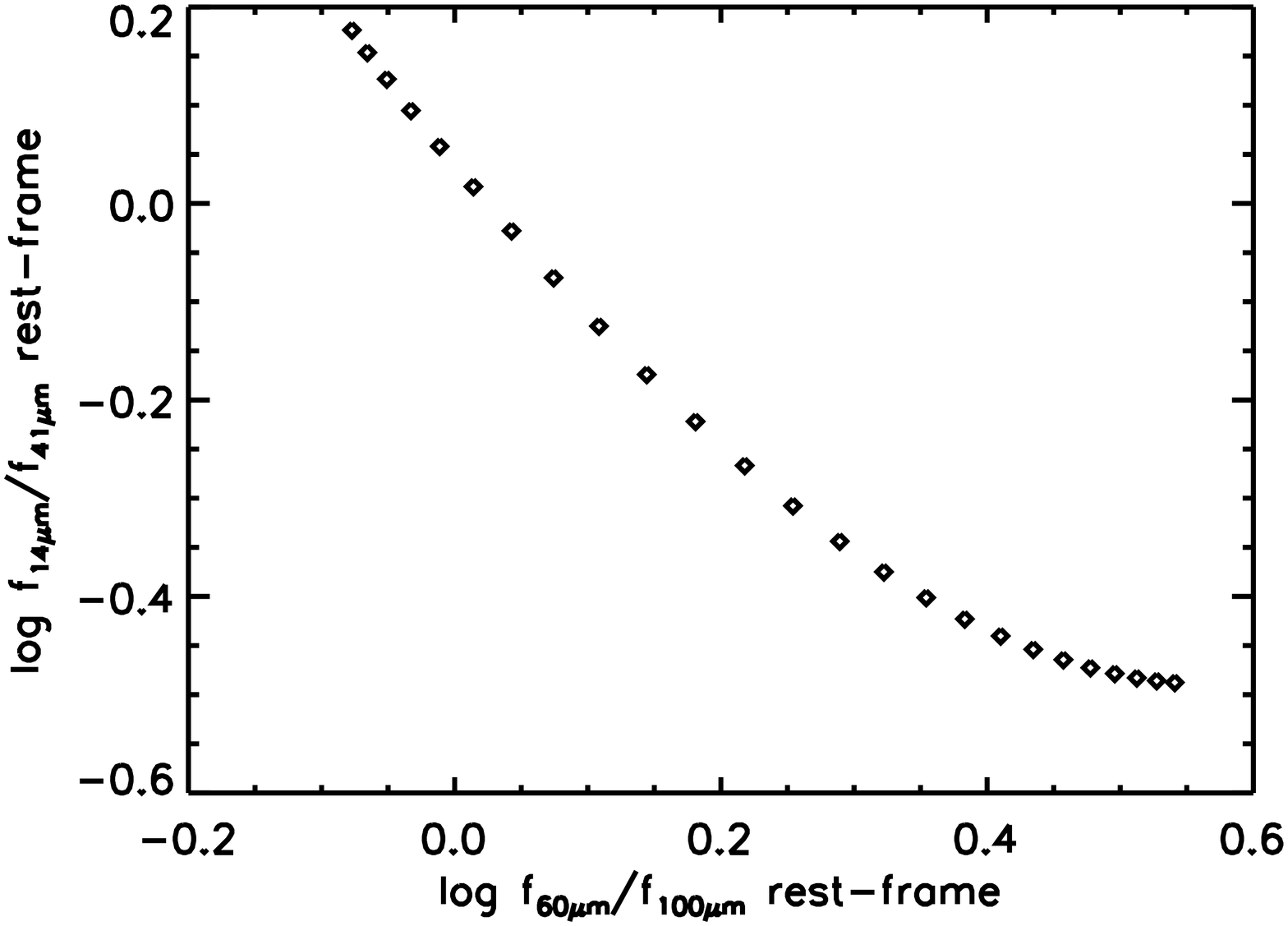}
     \caption{ Variations o{f $f_{14\mu m}$/$f_{41 \mu m}$ versus $f_{60\mu m}$/$f_{100 \mu m}$ (expressed in Jansky) for the DH02 templates used in the paper, corresponding to the  $\alpha$ range [1.0 - 2.5].} }
      \label{DH_CE}
\end{figure}

\begin{figure*} 
\centering
 \includegraphics[width=16cm]{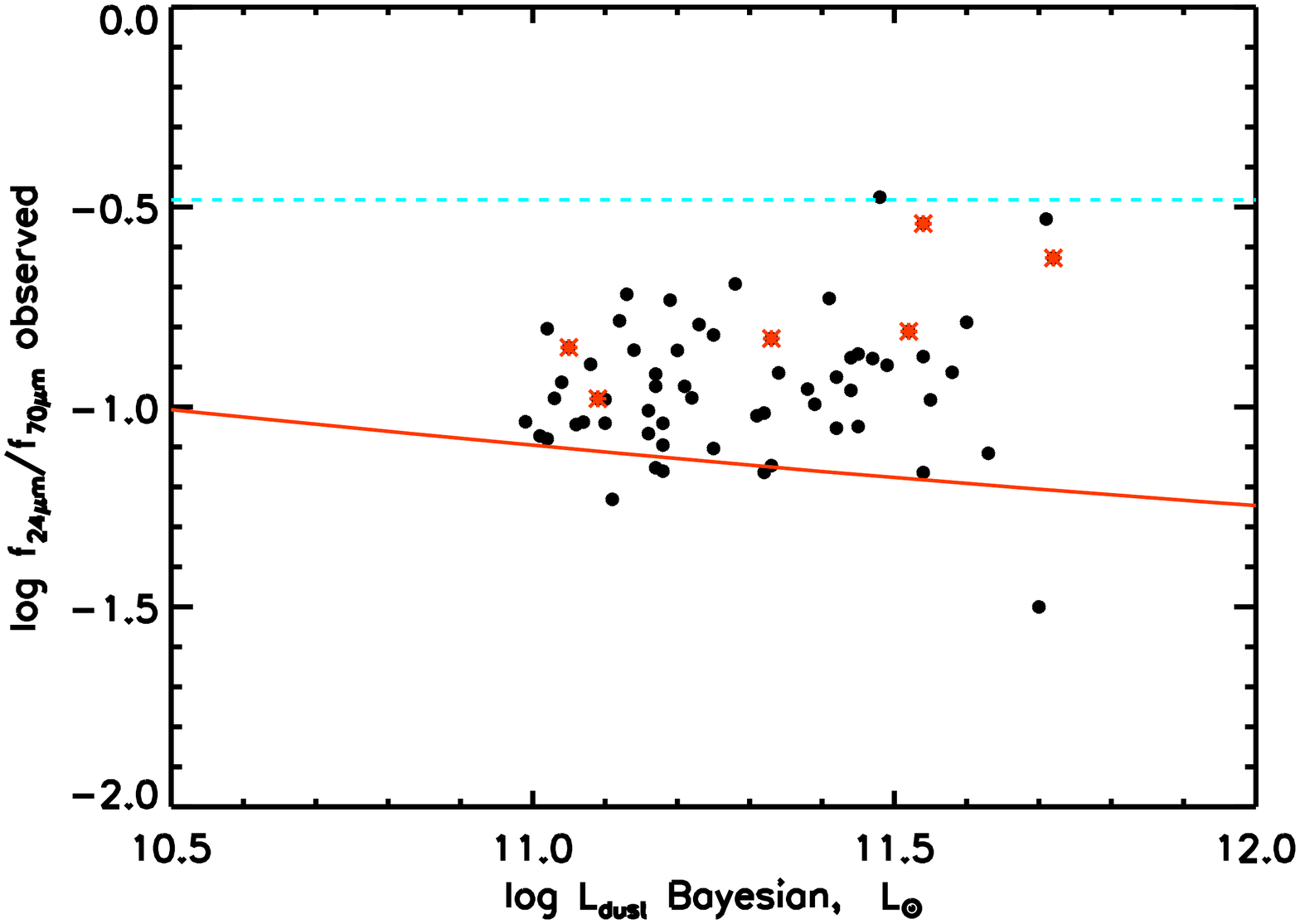}
     \caption{ This figure illustrates the ratio between the observed fluxes at 24 $\mu$m  and 70 $\mu $m as a function of the $L_{dust}$ estimated by the code. The solid line corresponds to the results given by the DH02 templates at z=0.7. 
 The dashed cyan line is given by \citet{siebenmorgen04a} for typical AGN contaminated object. 
 AGNs described in section 4.2 are shown as red stars.
     }
      \label{LIR_L24L70}
\end{figure*}

{We use the fluxes ratio $f_{24\mu m}$/$f_{70 \mu m}$ (corresponding to the highest wavelengths in our sample) in order to estimate the dust temperature.
The  observed $f_{24\mu m}$/$f_{70 \mu m}$ corresponds to  $f_{14\mu m}$/$f_{41\mu m}$ rest-frame.  }
{All our discussions are made in the framework of the DH02 models in the  [1.0-2.5]  $\alpha$ regime (see. Table \ref{table4}).}
{$f_{14\mu m}$/$f_{41\mu m}$ evolves in the opposite way of $f_{60\mu m}$/$f_{100\mu m}$ (i.e. $T_{dust}$, see Fig. \ref{DH_CE}). } 

Fig. \ref{LIR_L24L70} exhibits the observed $f_{24\mu m}$/$f_{70 \mu m}$ ratio as a function of $L_{dust}$ obtained for the LIRGs in the   70 $\mu $m sample.
{Predictions of DH02 models are overplotted. They are calibrated with the Marcillac et al. (2006) local relation between $f_{60\mu m}$/$f_{100\mu m}$ (i.e. $\alpha$) and $L_{dust}$ }. For a given $L_{dust}$ we find that our galaxies have a $f_{14\mu m}$/$f_{41\mu m}$ ratio higher than predicted by the models. This implies that {they have a lower $f_{60\mu m}$/$f_{100\mu m}$ ratio and thus are colder than expected.} We cannot attribute this result to the presence of AGN contaminated objects regarding the small fraction of flagged sources. 


\begin{figure*} 
\centering
\includegraphics[width=16cm]{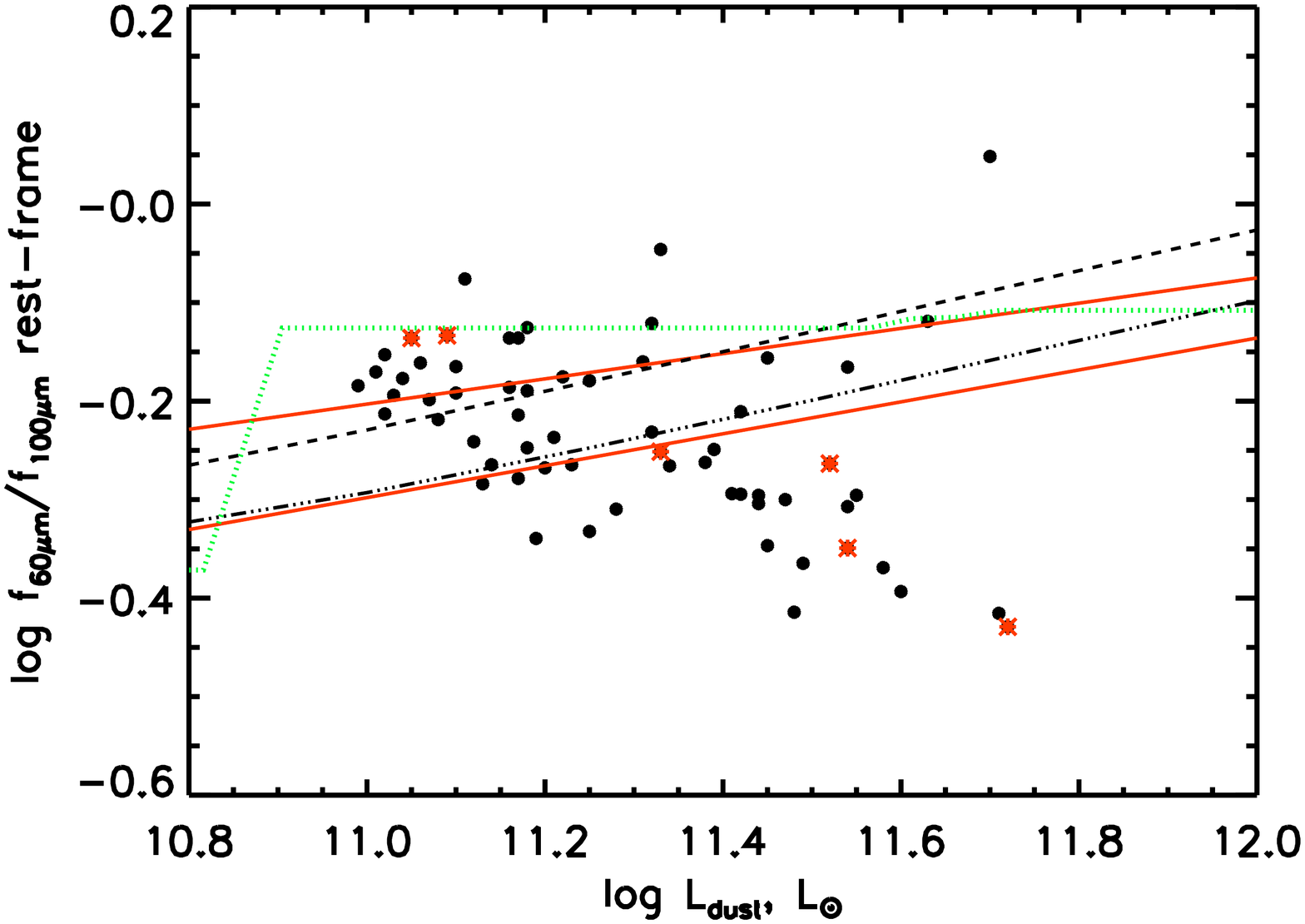}
     \caption{Comparaison of  the relation  log R(60/100) - log$_{10}L_{dust}$  given by IR SEDs of DH02 after the fit to the data, with the values given by local relations and models. 
     Upper and lower solid red lines represent local relations of  \citet{chapman03} and \citet{marcillac06}, respectively. We overplot models of CE01 (green dotted line), \citet{chapin09} (dashed black line) and \citet{valiante09} (dash-dotted black line).
     AGN detections are shown as red stars.}
      \label{R60_100}
\end{figure*}

{ We can also directly use the $f_{60\mu m}$/$f_{100\mu m}$ ratio deduced from the $\alpha$ parameter estimated by CIGALE.}
{In Fig. \ref{R60_100} we compare $L_{dust}$ and the derived $f_{60\mu m}$/$f_{100\mu m}$ values of our 70 $\mu$m  sample.
}It will have no sense to make this comparison for the rest of the sample for which we estimated $\alpha$  with the median value of  $\alpha$ found for the sub-sample. 
 We add to this plot the local  $L_{dust} - $$f_{60\mu m}$/$f_{100\mu m}$ relations of \citet{chapman03} and \citet{marcillac06}, and the luminosity-temperature models proposed by CE01, \citet{chapin09}, and  \citet{valiante09} in the luminosity range [$10^{11} ; 10^{12}$] L$_\odot$.    
 Fig.\ref{R60_100} shows that the code gives a different trend for the relation $f_{60\mu m}$/$f_{100\mu m}$  versus $L_{dust}$ than both local relations and models. Galaxies having the lowest IR luminosity are in agreement with them, but more luminous objects are found  colder than predicted, and globally they evolve in the opposite way {to local relations and models}.  
\citet{chanial07} make a selection at 60 and 170 $\mu $m using respectively $IRAS$ and $ISO$ observations at z=0 and find that galaxies selected at 170 $\mu $m are colder than ones detected at {shorter wavelengths}. They conclude that selecting galaxies at long wavelengths introduce a bias towards cold galaxies. 
Selecting the sample at a wavelength longwards of the peak, by definition, does not predispose towards warm sources (\citet{sandersmirabel96}). 
 Our sample is selected by rest-frame $\sim $ 41 $\mu $m emission, i.e. on warm dust; yet, we mostly find colder $T_{dust}$ at a given luminosity than predicted by models of DH02 and CE01, calibrated in the nearby universe. 
This tends to support the interpretation that the offset found towards colder temperatures at a given luminosity is at least in part a real difference.
\citet{zheng07} suggest that this tendency towards colder  $T_{dust}$  reflects a difference in dust and star formation geometry: whereas local LIRGs tend to be interacting systems with relatively compact very intense star formation and of comparable masses (e.g., \citet{wang06}), 0.5$<$z$<$1 LIRGs tend to be disk-dominated, relatively undisturbed galaxies (\citet{zheng04}; \citet{bell05}; \citet{melbourne05}). They suggest that these disk-galaxies host widespread intense star formation, like star formation in local spirals but scaled up, leading to relatively cold $T_{dust}$.

Our results are in agreement with  \citet{symeonidis09}. They compare FIR properties of LIRGs and Ultra Luminous IfraRed Galaxies (ULIRGs, $L_{dust} > 10^{12} $L$_{\odot}$) in the 0.1$<$z$<$2 redshift range, based on MIPS and IRAC data with $L_{dust} > 10^{10} $L$_{\odot}$,   70 $\mu $m selected objects,  with those of local sources of equivalent luminosity from the local $IRAS$ Bright Galaxy Sample. 
 They fit their data with the  \citet{siebenkrugel07} theoretical SED templates and find that their distant LIRGs and ULIRGs are colder than their local equivalents. 
They found a significant cold-dust associated FIR excess not detected in local sources of comparable luminosity directly linked to the evolution in dust and star formation properties from the local to the high redshift Universe.
 
{$Herschel$ is now providing} longward 70 $\mu $m data allowing a reliable estimation of  $T_{dust}$.


 \subsection{ Star formation rate and stellar mass}

 \begin{figure*} 
 \sloppy
\centering
 \includegraphics[width=17cm]{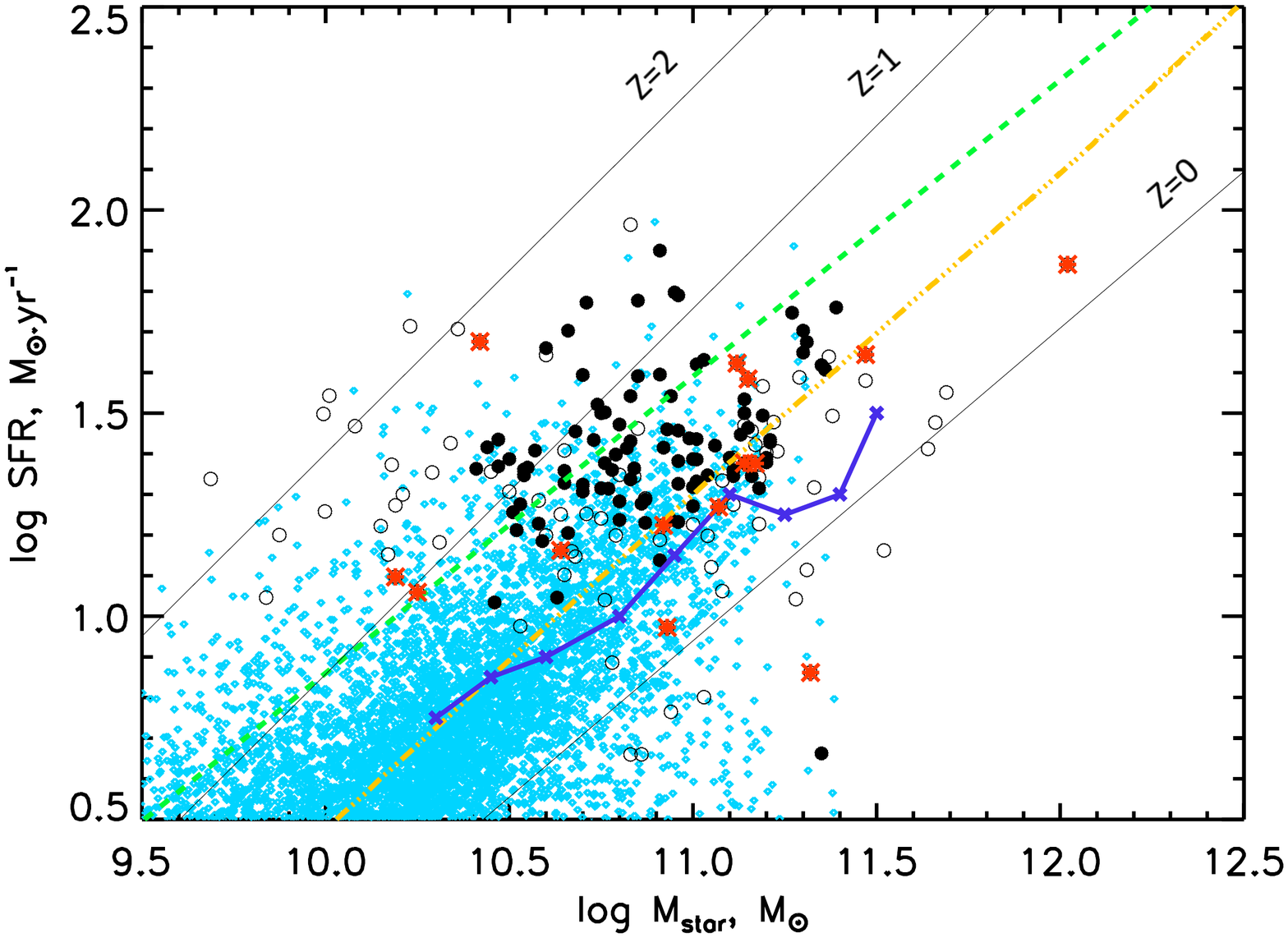}
     \caption{The relation SFR-$M_{\star}$ for the total sample. Filled circles (empty circles) illustrate galaxies for which the age of the young stellar population is found greater (lower or equal) than 0.3 Gyr by CIGALE. AGNs are represented by red stars.
 We overplot  the semi-analytical Millennium simulations (cyan diamonds), analytical models of \citet{buat08} (dash-dotted yellow line) at z=0.7 and \citet{noeske07a} (connected blue crosses) at z$\sim$0.7, and  observations of \citet{elbaz07} (solid black line, z=0 and 1), \citet{daddi07} (solid black line z=2) and \citet{santini09} at z$\sim$0.7 (dashed green line).
 }
      \label{SFR_Mstar}
\fussy
\end{figure*}

\begin{figure*} 
\centering
 \includegraphics[width=15cm]{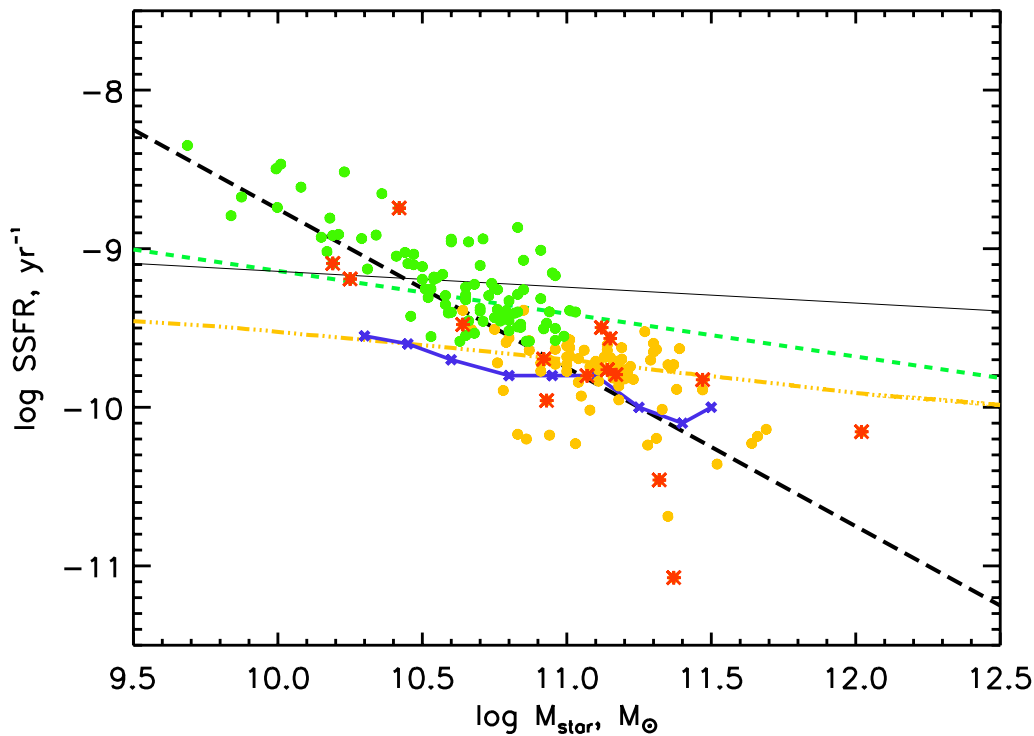}
    \caption{The SSFR-$M_{\star}$ relation for the total sample. Green dots and yellow dots represent objects for which $f_{ySP}$ are larger or lower than 10$\%$, respectively. AGNs contaminated objects are represented by red stars. Black dashed line shows the detection limit.
    We overplot analytical models of \citet{buat08} (dash-dotted yellow line) at z=0.7 and \citet{noeske07a} (connected blue crosses) at z$\sim$0.7, and  observations of \citet{elbaz07} (solid black line, z=1) and \citet{santini09} at z$\sim$0.7 (dashed green line).}
      \label{SSFR_Mstar}
\end{figure*}

How and when galaxies build up their $M_{\star}$ is still a major question in observational cosmology. While a general consensus has been reached in recent years on the evolution of the galaxy stellar mass function (e.g., \citet{Dickinson03}; \citet{drory04}; \citet{bundy05}; \citet{pannella06}; \citet{marchesini08}), the redshift evolution of the SFR as a function of $M_{\star}$ still remains unclear. 
The evolution of the galaxy SFR-$M_{\star}$ relationship provides key constraints on $M_{\star}$ assembly histories of galaxies. 
For star-forming galaxies a correlation is observed between SFR and $M_{\star}$ with a slope near to unity  from z $\sim$ 0 to 2 and a scatter of 0.3 dex (\citet{daddi07}; \citet{elbaz07}; \citet{noeske07a}). 
The slope and scatter are expected to be a direct reflection of the similarity in the shape of galaxy SFHs among various models and at various masses (\citet{dave08}, \citet{noeske07b}).
It is worth noting that other z = 2 observations suggest a somewhat large scatter (\citet{shapley05}; \citet{papovich06}), so the observational picture is not entirely settled.


Fig.\ref{SFR_Mstar} shows the SFR of each galaxy of our sample as a function of the corresponding $M_{\star}$, 
and the results of various models and observations :  analytical models of \citet{buat08} and \citet{noeske07a}, the semi-analytical Millennium simulations of  \citet{KW07} used in \citet{buat08} and  observations of \citet{elbaz07}, \citet{daddi07}, and \citet{santini09}.
Our sample of LIRGs has an average redshift of $\sim$ 0.7. We find the vast majority of the sample between the correlations at z=0 and z=1 given by \citet{elbaz07}. 
Our data cover the extreme part of the Millennium simulations where we can find galaxies with the largest SFRs between $\sim10-100$ $\mathrm{M}_{\odot} \mathrm{yr}^{-1}$.
We do not observe a correlation between SFR and $M_{\star}$, but instead a flat distribution of SFR. 
We consider a tight range of SFR since we select objects with  $L_{dust}$ $\geqslant$$10^{11}$L$_\odot$ (see below) : this range is not sufficient to search for a general correlation between $M_{\star}$ and the SFR.
However in the considered SFR range we observe a large scatter in $M_{\star}$.
Indeed galaxies whose young stellar population is younger than 300 Myr exhibit a larger dispersion in $M_{\star}$ than galaxies whose SFR is constant over a longer period (Fig.\ref{SFR_Mstar}).
Therefore the SFH of our objects has an influence on the dispersion of the SFR-$M_{\star}$ plot. This is in agreement with the conclusions of \citet{dave08} and \citet{noeske07b}.


The specific star formation rate (SSFR) defined as the ratio between SFR and $M_{\star}$ (e.g., Kennicutt et al. (2005)) is also commonly used to trace the SFH in galaxies. 
The SSFR is plotted in Fig.\ref{SSFR_Mstar} as a function of the corresponding $M_{\star}$ and we can see the increase of the SSFRs towards lower $M_{\star}$.
Before any interpretation we can estimate a detection limit in SSFR : $L_{dust}$  = $10^{11}$  L$_{\odot}$ translates into SFR$_{lim}$ = 17.78 $\mathrm{M}_{\odot} \mathrm{yr}^{-1}$ at z = 0.7 ( Eq. (1) of Buat et al. (2008)).
We overplay this detection limit and see that the trend of SSFR with $M_{\star}$ is largely constrained by this limit.
We compare our results with the same models and observations as illustrated in Fig.\ref{SFR_Mstar}. 
SSFRs of our intermediate-mass galaxies ($M_{\star}$$\leq $$10^{11} $M$_{\odot}$) are in agreement with those observed by \citet{santini09}, but for these values of $M_{\star}$ we cannot detect more quiescent galaxies because of the detection limit.
SSFRs of large-mass galaxies ($M_{\star}$$>$ $10^{11}  $M$_{\odot}$) are better described by the models of \citet{buat08}. 
 These massive galaxies have SSFRs which do not exceed 2.5 $10^{-10}$yr$^{-1}$ (corresponding to a SFR$\sim$25 $\mathrm{M}_{\odot} \mathrm{yr}^{-1}$) and never reach the SSFRs of intermediate-mass galaxies observed in this study.
  In order to distinguish galaxies for  which the young stellar population significantly contributes to $M_{\star}$, we identify objects for which $f_{ySP}$ estimated with CIGALE is larger or lower than 10$\%$ and see that the 2 groups are well-separated according to their stellar mass:  galaxies with $M_{\star}$ $\leq$ $10^{11}  $M$_{\odot}$ have more than 10$\%$ of their mass due to the young stellar population, whereas in galaxies with $M_{\star}$ $>$ $10^{11}  $M$_{\odot}$ this mass represents less than 10$\%$. 
This result is in agreement with the common picture that massive galaxies formed the bulk of their stars early and on shorter timescales, while numerous less massive galaxies evolve on longer time-scales, a phenomenon generally linked to the ÒdownsizingÓ scenario (e.g. Cowie et al. (1996), \citet{BE00} ; \citet{fontana03} ; \citet{feulner05} ; \citet{perez-gonzalez05} ; \citet{papovich06} ; \citet{damen09}).


\section{Conclusions}
We  performed a SED-fitting analysis of a sample of LIRGs at z=0.7 detected at 24 $\mu $m by $Spitzer$/MIPS and for which 80 $\%$ of the objects have a detection at 231 nm by $GALEX$. 
This sample is observed over a large range of wavelengths, from NUV to FIR (70 $\mu $m) and is made of 181 galaxies which 62 are detected at 70 $\mu $m.
We fit the SEDs of our galaxies with the CIGALE code (\citet{burgarella05}, \citet{noll09}) which combines stellar and dust emissions in a physical way.  This study is the first use of CIGALE at redshift larger than 0.
The stellar populations synthesis code of   \citet{maraston05} is adopted to model the stellar emission (UV, optical and NIR wavelengths). Then the code uses semi-empirical one-parameter models of \citet{DH02} to reproduce the dust emission in MIR-to-FIR wavelengths.

CIGALE allows the estimation of several physical parameters based on a Bayesian-like analysis.
We show that we are able to estimate reliable stellar masses, star formation rates, fractions of young stellar population,  \citet{DH02}  IR-templates parameter $\alpha$ (if 70 $\mu $m is available), dust attenuation, dust luminosities as well as the fraction of dust emission due to an active galactic nucleus. The investigation is performed by building mock catalogues based on the real samples.

The 70 $\mu $m sample exhibits colder dust temperatures (as traced by the ratio $L_{24\mu m}/L_{70\mu m}$ and the $\alpha$ parameter of the \citet{DH02} templates) than expected from local relations between dust luminosity and temperature, confirming other recent results for similar samples. 

Our LIRGs appear to form stars actively; they exhibit a flat distribution with a large scatter in the star formation rate - stellar mass plot.
The amplitude of the dispersion is related to the age of the young stellar population, a tighter distribution being found for the largest ages.

We find that our galaxies with a stellar mass $>$ $10^{11} $M$_{\odot}$ have less than 10$\%$ of their mass coming from the young stellar population. The specific star formation rate for these massive  galaxies never reaches the one found for intermediate-mass galaxies of our sample, confirming the downsizing scenario. 

The multiwavelengths data analysis performed in this study provide reliable estimates of several physical parameters but may turn out insufficient to determine accurate dust temperatures. Forthcoming data from $Herschel$ will help us to better constrain galaxies SEDs  and thus to derive more reliable parameters.

\end{document}